\documentclass[12pt]{article}  

\usepackage{graphicx}
\usepackage{amssymb}  

\begin{document}

\title{\bf Quark matter in compact stars?}


\author{Alford,~M.$^{1}$,\hspace{0.5em}
Blaschke,~D.$^{2,3}$,\hspace{0.5em}
Drago,~A.$^{4,5}$, \hspace{0.5em}
Kl\"ahn,~T.$^{3,6}$,\\
Pagliara,~G.,$^{4,5}$,\hspace{0.5em}
Schaffner-Bielich,~J.$^{7}$
}

\date{Dec 25, 2006}

\maketitle

\begin{enumerate}
\item Department of Physics, Washington University, St Louis, MO 63130, USA
\item Instytut Fizyki Teoretycznej, Uniwersytet Wroclawski, pl. M. Borna 9,
50-204 Wroclaw, Poland
\item Institut f\"ur Physik, Universit\"at Rostock, D-18051 Rostock, Germany
\item Dipartimento di Fisica, Universit{\`a} di Ferrara, I-44100 Ferrara, Italy
\item INFN, Sezione di Ferrara, I-44100 Ferrara, Italy
\item Gesellschaft f\"ur Schwerionenforschung mbH (GSI), 
  Planckstr.~1, 64291 Darmstadt, Germany
\item Institut f\"ur Theoretische Physik, Goethe Universit\"at, 
D-60438 Frankfurt am Main, Germany
\end{enumerate}

In a theoretical interpretation of observational data from the
neutron star EXO 0748-676, \"Ozel
concluded that quark matter probably does not exist in the center of
neutron stars~\cite{O2006}. However, this conclusion was based on a 
limited set of possible equations of state (EoS) for quark matter.
Here we compare \"Ozel's observational limits with predictions
based on a more comprehensive set of proposed quark matter equations
of state from the existing literature, and conclude that the presence
of quark matter in EXO 0748-676 is not ruled out.

\"Ozel's stated lower limits on the mass and radius are
$M\geqslant 2.1\pm 0.28~M_\odot$ and 
$R\geqslant 13.8\pm 1.8~{\rm km}$. She correctly
points out that these values exclude a soft EoS. She then
infers that there is no quark matter in this compact star. However, this 
conclusion does not follow because
quark matter can be as stiff as nuclear matter, as effects from
strong interactions (QCD) can harden the EoS
substantially. The corresponding hybrid or quark stars can indeed reach
a mass of $2M_\odot$ as demonstrated in calculations
using the MIT bag model~\cite{ABPR2005}, perturbative corrections to
QCD~\cite{FPS2001}, and the Nambu--Jona-Lasinio
model~\cite{Ruster:2003zh}. The mass-radius relations for compact stars
using various quark matter and nuclear matter EoS along
with the lower limits derived by \"Ozel are shown in Figure 1. 

In addition to the mass and radius, there are potential constraints on
(or signatures of) the presence of quark matter from observations of
the cooling, spin-down, and precession of neutron stars, and from
transient phenomena such as glitches, magnetar flares, and
superbursts.

Cooling observations of firmly-identified neutron stars
are mostly consistent with a ``minimal model'' of
nuclear matter cooling, but there is evidence of faster cooling
in limits obtained from supernova remnants, and the presence of exotic
forms of matter is not ruled out~\cite{Page:2005fq}.
A detailed analysis of cooling data
including information from elliptic flow in heavy 
ion collisions was unable to find any purely nuclear EoS that was compatible
with all the data~\cite{K2006}. Models involving some quark matter in the
cores of neutron stars were more successful~\cite{GBV2005,Klahn:2006iw}.

Measurements of the spin-down rate of neutron stars can be used
to constrain the shear and bulk viscosity of the interior, since
sufficiently low viscosity would lead to very fast spin-down via
gravitational radiation from unstable r-modes. Preliminary
calculations rule out a strange star made of CFL matter \cite{Madsen:1999ci},
but hybrid stars are not ruled out. More controversially,
it has also been argued that
the measured precession of some stars is
inconsistent with the standard understanding of nuclear matter 
\cite{Link:2003hq}.


Glitches (temporary speeding-up in the rotation of a neutron star
that is gradually spinning down) are only partially understood, but
are believed to provide evidence for a substantial crust
overlapping with a superfluid region inside the
star \cite{glitch}. This does not exclude the presence of a 
quark matter core, and may not even exclude strange stars, since
there are superfluid and crystalline phases of quark matter 
\cite{CFL,Alford:2000ze}.

Observations of quasi-periodic oscillations in soft
gamma repeaters have recently been used to obtain the frequencies
of toroidal shear modes of their crusts. The results are not consistent
with these objects being purely strange stars, but put no limits on the
presence of a quark matter core inside them \cite{Watts:2006hk}.

Finally, observations of superbursts in low-mass X-ray binaries
yield ignition depths much smaller than those predicted for
standard neutron stars, and are more compatible with these
objects being hybrid stars with a relatively thin baryonic crust
\cite{Page:2005ky}.

We conclude that if \"Ozel's analysis is correct, it can be used to put
constraints on the parameters of the quark matter EoS, but neither
Ozel's analysis nor the other available observational data
has yet ruled out the presence of deconfined quarks in compact stars.

\begin{figure}
\begin{center}
\includegraphics[width=0.9\hsize]{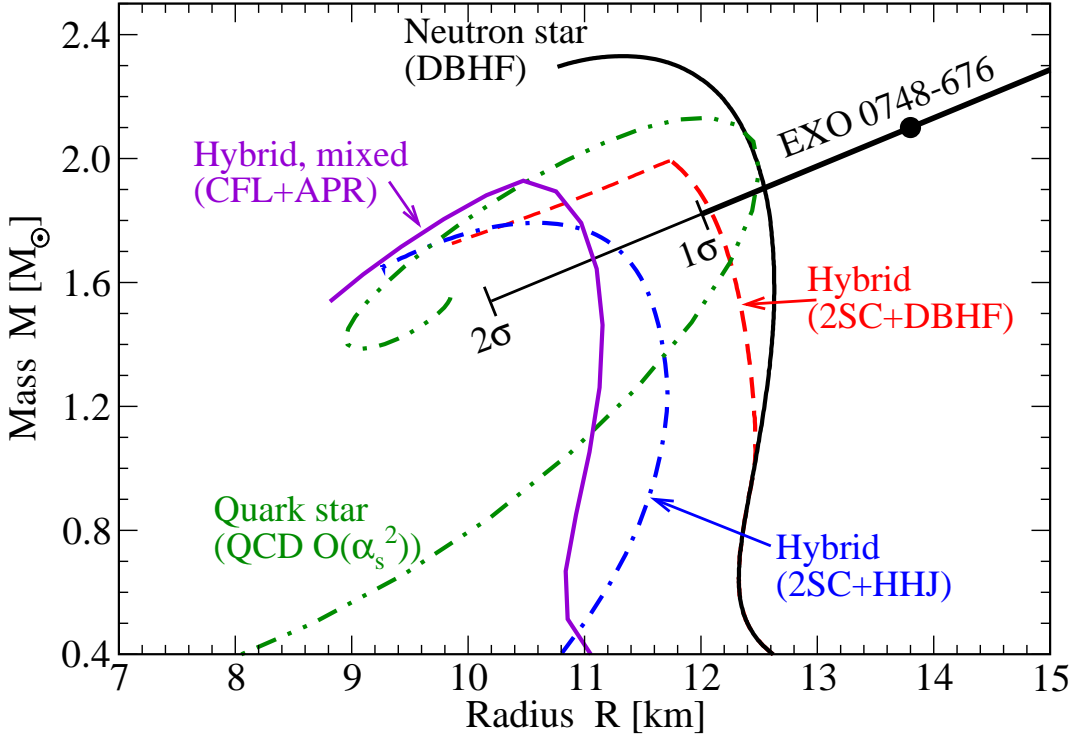}
\caption{The lower limit on $M$ and $R$ from \"Ozel's analysis of
  EXO~0748-676 \cite{O2006} (we show one and two sigma error bars), and
  the calculated $M$-$R$ curve for various quark matter and nuclear
  matter equations of state. These include pure nuclear matter described
  by the DBHF (relativistic Dirac-Brueckner-Hartree-Fock)
  EoS \cite{K2006}; a hybrid star
  with a core of 2SC quark matter (a two-flavor color-superconducting
  phase in which only the up and down quarks form Cooper pairs) 
  and a mantle of DBHF nuclear matter \cite{Klahn:2006iw}; a
  hybrid star with a core of 2SC quark matter and mantle of HHJ nuclear
  matter (APR with high-density causality corrections) 
  \cite{GBV2005}; a hybrid star whose core is a mixed phase of APR
  nuclear matter (based on the Argonne $v_{18}$ two-nucleon interaction
  with variational chain summation) 
  and CFL quark matter
  (the ``color-flavor-locked'' color-superconducting phase in which all
  three colors and flavors undergo Cooper pairing) 
  \cite{ABPR2005}; and a pure quark matter star using an equation of
  state with ${\cal O}(\alpha_s^2)$ QCD corrections \cite{FPS2001}.  It
  is clear that the presence of quark matter is not excluded by the
  EXO~0748-676 results.  }
\end{center}
\label{fig:MR}
\end{figure}

\end{document}